\documentstyle[11pt,A4WIDE]{article}

\begin{document}
\newcommand{\be}{\begin{equation}}
\newcommand{\ee}{\end{equation}}
\newcommand{\nue}{\bar \nu_{e}}
\newcommand{\MeV}{\rm MeV}
\def\blankline{\par \vspace{\baselineskip}\noindent}
\def\journalfont{\it}         
\def\jou#1{{\journalfont #1\ }}
\def\nat{\jou{  Nature}}
\def\ncim{\jou{ Nuovo Cim.}}
\def\nucp{\jou{ Nuc.\ Phys.}}
\def\ncb{\jou{  Il Nuovo Cimento ``B}}
\def\pl{\jou{   Phys.\ Lett.}}
\def\pr{\jou{   Phys.\ Rev.}}
\def\prep{\jou{ Phys.\ Rep.}}
\def\prl{\jou{  Phys.\ Rev.\ Lett.}}
\def\ptp{\jou{  Prog.\ Theor.\ Phys.}}
\def\rmp{\jou{  Rev. Mod. Phys.}}
\def\spj{\jou{  Sov.\ Phys.\ JETP}}
\def\np{\jou{  Nucl.\ Phys.}}
\begin{titlepage}
\begin{flushright}
DFRM1 n.944, 21 May '93\\
DFPD 93/TH/36\\
SISSA AP/93/65\\
\end{flushright}
\vspace{24pt}
\centerline{\bf \Large Electromagnetic properties of the neutrinos and
the pion radiative decay }
\vspace{24pt}
\begin{center}
{\large D. Grasso$^{a,}$\footnote{Email: grasso@roma1.infn.it}
, M. Pietroni$^{b,c,}$\footnote{Email: pietroni@mvxpd5.pd.infn.it}
 and A. Riotto$^{c,d,}$\footnote{Email:riotto@tsmi19.sissa.it}}
\end{center}
\vskip 0.2 cm
{
\vskip 0.2 cm
\centerline{\it $^{(a)}$ Istituto Nazionale di Fisica Nucleare,}
\centerline{\it Sezione di Roma, Dipartimento di Fisica, Piazzale A.
Moro, 2 - 00185 Roma, Italy}
\vskip 0.2 cm
\centerline{\it $^{(b)}$Dipartimento di Fisica Universit\`a di Padova,}
\centerline{\it Via Marzolo 8, 35100 Padua, Italy.}
\vskip 0.2 cm
\centerline{\it $^{(c)}$Istituto Nazionale di Fisica Nucleare,}
\centerline{\it Sezione di Padova, 35100 Padua, Italy.}
\vskip 0.2 cm
\centerline{\it $^{(d)}$International School for Advanced Studies,
SISSA-ISAS}
\centerline{\it Strada Costiera 11, 34014 Miramare, Trieste, Italy}}
\vskip 0.5 cm
\centerline{\large\bf Abstract}
\vskip 0.2 cm
\baselineskip=24pt
We investigate the implications of neutrino electromagnetic dipole
moments for the radiative decay $\pi \rightarrow e
\nu \gamma$.
We show that the dominant new effect  comes from the interference between
the amplitude of the process with the photon emitted by the neutrino
and the structure dependent standard amplitude.
 Such interference takes place only if the massive neutrino is a Majorana
fermion.  Moreover, since electric and magnetic dipole moments
give rise to contributions with opposite signs, future experiments
might be able to distinguish between them.
\end{titlepage}
\baselineskip=20pt

\section{Introduction}

The study of the pion decay has represented historically one of the most
important resources to investigate the structure of the weak interactions.
Although the dominant two body decay modes are well understood in the
framework of the standard model (SM) some disagreements with the theory
seems to appear for the radiative decay modes. The only experiment probing
the mode $\pi^+ \rightarrow \mu^+ \nu_{\mu} \gamma$ showed an excess of
events in the high energy range of the outgoing muon \cite{Cast}.
Hopefully, a definitive answer to this issue will be given in a
planning experiment \cite{rapid}.
More recently a measurement has been performed for the rate of the radiative
decay in flight
$\pi^- \rightarrow e \nue \gamma$ in a wide range of the kinematical variables
\cite{Bolotov}. The authors of this work claim for a deficit of events
( more than three standard deviations less than the SM prediction) in
the ${SD}^-$ (see below) `structure dependent' component of the spectrum.
This effect
has been interpreted by Poblaguev \cite{Pobla} as a possible signal of a new
tensorial quark-lepton interaction. Even if this kind of coupling could be
induced by
supersymmetric effects it was shown that their strength is two orders of
magnitude  smaller than the needed value \cite{Bel}.
Moreover, Voloshin has pointed out that such a tensorial coupling is strongly
constrained by the rate of the process $\pi \rightarrow e\nue$,
which suggests to focus on the leptonic sector \cite{Volosh}.

In this letter we investigate the implications for the radiative
decay $\pi \rightarrow e \nu_e \gamma$ of hypothetical neutrino
electromagnetic dipole moments (DM's).
Although any electromagnetic interaction of the neutrinos is forbidden
by the SM several classes of models foresee non-vanishing magnetic or electric
DM. Furthermore, the indication of a correlation of solar
activity with solar neutrino flux could find an elegant explanation
if the electronic neutrino had a magnetic (or electric) DM of about
$10^{-11} \mu_B$ \cite{VVO}.

The present experimental limits on the magnetic ($\mu$) and the electric
($d$) DM's  are \cite{limits} \cite{grotch}
\be
\left|\mu_{\nu_e} + id_{\nu_e}\right| < 1.5 \times 10^{-10}\:\mu_B\:\:\:\:,
\left|\mu_{\nu_{\mu}} + id_{\nu_{\mu}}\right| < 1.2 \times 10^{-9}\:\mu_B
\:\:\:\:,
\left|\mu_{\nu_{\tau}} + id_{\nu_{\tau}}\right|
< 4 \times 10^{-6} \mu_B \label{limiti}
\ee
where  $\mu_B=e^2/2m_e$ is the Bohr magneton. Slightly
more stringent bounds for neutrinos lighter than about 1 MeV come from
astrophysical considerations \cite{raffelt1}.

We are mainly interested in non diagonal DM's since, as we shall see,
the main non-standard contribution to the  radiative decay rate
$\pi\rightarrow e \nue \gamma$ might come from a DM mixing
the first and the third neutrino mass eigenstates.
The same limits set for $\mu_{33}$ or $d_{33}$, see eq. (\ref{limiti}),  also
apply to $\mu_{13}$
and $d_{13}$, since the experimental analysis \cite{grotch}
does not discriminate between the neutrino families.
Moreover, $\nue e$ scattering experiments, which provide the limits on
$\mu_{11}$ ($d_{11}$) and $\mu_{12}$ ($d_{12}$), are performed at
energies of the order of 1 MeV, thus they do not constrain $\mu_{13}$
($d_{13}$) if the $\tau$ neutrino mass is not far from its upper
experimental bound $m_{\nu_3} < 35$ MeV.

\section{Interference of DM with conventional amplitudes}

The amplitude of the radiative decay $\pi^- \rightarrow e \nue \gamma$
in the framework of the SM is \cite{Bryman}:
\be
{\cal M}_{\pi\rightarrow e\nue \gamma} = {\cal M}_{IB} + {\cal M}_{SD}
\ee
where
\be
{\cal M}_{IB} = i\,{e G_F f_{\pi} \cos\theta_c m_e\over \sqrt{2}}\,{\bar u_e}
\biggl( {1\over 2}\, { {\hat \epsilon}{\hat k} \over pk} + {p\epsilon \over pk}
- {q\epsilon \over qk} \biggr) (1 - \gamma_5)\, v_{\nue} \label{MIB}\ ,
\ee
\be
{\cal M}_{SD} = {e G_F m_{\pi}^2 \cos\theta_c \over \sqrt{2}}\,{\bar u_e}
\gamma^{\alpha} (1 - \gamma_5)\, v_{\nue} \bigl[ F^V\,
\varepsilon_{\alpha\beta\sigma\tau}q^{\sigma}k^{\tau} + i\, F^A\,
(g_{\alpha\beta}\,qk - k_{\alpha}q_{\beta}) \bigr] \epsilon^{\beta *}
\label{MSD}
\ee
are the QED corrections to $\pi^- \rightarrow e\nue$ Inner Bremsstrahlung (IB)
and Structure Dependent (SD) amplitudes respectively.
In the equations (\ref{MIB}) and (\ref{MSD}) $\theta_c$ is the Cabibbo angle,
$f_{\pi} \simeq 130\: \MeV$ is the pion decay constant, $\epsilon^{\beta}$ is
the photon polarization, $F^{V\,(A)}$ is the vectorial (axial) pion
electromagnetic form-factor. The vectorial form-factor calculated from the
measured $\pi^0$ lifetime, using CVC, is $F^V = 0.0259\pm0.0005$, while
$F^A =(0.41 \pm 0.23)|F^V|$ from ref.
\cite{Bolotov}.

We now assume that the neutrinos are massive and have magnetic and/or electric
DM (see appendix A). In this case the outcoming photon can be attached
to the neutrino leg (see Fig. 1) via the matrix element in eq. (\ref{A1}).
The amplitude for this process is
\be
{\cal M}_{DM} = - i\,{G_F f_{\pi} \cos\theta_c\over \sqrt{2}} \sum_{ji}
U_{ei}^*U_{je} \,{\bar u_e}\,\hat{q}\,(1-\gamma_5)
{1\over \hat{k} +
\hat{h} - m_i + i\epsilon}\,(\mu_{ji} + i\,d_{ji}\gamma_5)
\sigma_{\alpha\beta}k^{\beta}\epsilon^{\alpha *}\,v_{\nu_e} \label{MDM}
\ee
where $U_{aj}$ are the mixing matrix elements between  weak ($\nu_a$) and
mass ($\nu_i$) neutrino eigenstates, defined according to $\nu_a =
\sum_{j=1}^3 U_{aj}\nu_j$.
The square of the total amplitude ${\cal M} ={ \cal M}_{IB} + {\cal M}_{SD} +
{\cal M}_{DM}$ summed over the photon polarizations and the spins of the
outgoing fermions is
\be
\overline{|{\cal M}|^2} = \overline {|{\cal M}_{IB}|^2} + \overline {|{\cal
M}_{SD}|^2} +
\overline{|{\cal M}_{DM}|^2} + 2 \overline {{\rm Re}({\cal M}_{IB}^{\dag}{\cal
M}_{SD})} +
2{\rm Re}(\overline{{\cal M}_{IB}^{\dag}{\cal M}_{DM}}) +
2 {\rm Re}(\overline{{\cal M}_{SD}^{\dag}{\cal M}_{SD}})\, .
\ee
We assume that the lighter neutrino eigenstate, ($\nu_{1}$), has a  vanishing
mass with respect to our energy scale ($m_{\pi}$). Then, the conventional
terms $\overline{|{\cal M}_{IB}|^2},\  \overline{|{\cal M}_{SD}|^2}$ and
$\overline{ {\rm Re}({\cal M}_{IB}^{\dag}{\cal M}_{SD})}$ are essentially
the ones
predicted by the SM.  It is convenient to write them in the pion rest frame
and in terms of the kinematical variables
\be
x \equiv {2E_{\gamma}\over m_{\pi}}\qquad
y \equiv {2E_{e}\over m_{\pi}}\qquad z \equiv {2E_{\nu}\over m_{\pi}}
\ee
where energy conservation implies $x + y + z = 2$. Then, one gets
\cite{Bryman}:
\be
\overline{|{\cal M}_{IB}|^2} = \Bigl({e G_F f_{\pi} \cos\theta_c\,m_e\over
\sqrt{2}}
\Bigr)^2\,{8(1 - y)\bigl[(1 + (1 - x)^2\bigr]\over x^2(x + y - 1)}
\ee
\be
\overline{|{\cal M}_{SD}|^2} = \Bigl({e G_F f_{\pi} \cos\theta_c m_{\pi}\over
\sqrt{2}}\Bigr)^2\,2\bigl(|F^V + F^A|^2\,SD^+(x,y) + |F^V - F^A|^2\,SD^-(x,y)
\bigr)\ ,
\ee
where $SD^+(x,y) = (1 - x)(x + y - 1)^2$ and $SD^-(x,y) = (1 - x)(1 - y)^2$,
and
\be
{\rm Re}(\overline{{\cal M}_{IB}^{\dag}{\cal M}_{SD}}) = \Bigl({e G_F f_{\pi}
\cos\theta_c \over \sqrt{2}}\Bigr)^2\,m_{\pi} m_e \bigl((F^V + F^A)F(x,y) +
(F^V - F^A)G(x,y)\bigr)
\ee
where
\be
F(x,y) = - {(1 - y)(1 - x)\over x}\qquad G(x,y) = {1 - y\over x^2}\Bigl(
1 - x + {x^2\over x + y - 1}\Bigr)\ .
\ee
This last term is negligible with respect to the previous ones in all the
permitted kinematical region, and we discard it hereafter.

A more careful discussion is worthful for the interference of the
electromagnetic DM's mediated decay amplitude (${\cal M}_{DM}$) with
${\cal M}_{IB}$ and ${\cal M}_{SD}$. The occurrence of the interference depends
on the different phases
of the electric and magnetic DM for Dirac and Majorana neutrinos.
It is already known that the electromagnetic properties of Majorana
neutrinos (if any) can be well different with respect to Dirac ones
\cite{Kayser}\cite{Petcov}.
In the present case the  interference with the conventional amplitude
cannot take place at all
for Dirac neutrinos, if CPT and CP are conserved, whereas it can
for Majorana ones.
We can understand it in a straightforward way by observing that the
imaginary part of the bubble diagram in  fig.2, {\it i.e.}
\begin{eqnarray}
{\rm Im} \, \rm{Tr}(A) = {(G_F f_{\pi}m_e)^2\over 2}
&{\rm Im}& \, \Biggl( \sum_{ji}
U_{ej}^*U_{ie}\, {\rm Tr}\Bigl(\hat{q}(1 - \gamma_5)\,iS_{\nu_j}(p)\,ik^{\nu}
\sigma_{\mu\nu}\,(\mu_{ji} + id_{ji}\gamma_5) \label{optic}\\
&\times&(-i)D^{\mu\rho}\,iS_{\nu_i}(p+k)
\,\hat{q}(1 - \gamma_5)\,iS_e(q-p-k)\,ie\gamma_{\rho}\,iS_e(q-p)\Biggr)
\nonumber
\end{eqnarray}
is non zero only if
\begin{equation}
{\rm Im}\:(\mu_{ij}+ i d_{ij}) \neq 0.
\label{imag}
\end{equation}
In eq. (\ref{optic}) $S_e,\;S_{\nu},\,D_{\mu\nu}$ are, respectively,
the real part of electron, neutrino, and photon propagators.
By the optical theorem we know that the amplitude in eq. (\ref{optic}) is
related to the decay rate for $\pi\rightarrow e\:\nu\:\gamma$, so that
the latter is zero unless the condition in eq. (\ref{imag}) is satisfied.
We show in appendix that this is impossible for Dirac neutrinos when CP is
conserved. On the other hand, for Majorana neutrinos the  electric and
magnetic DM's
carry a relative phase of $\pi/2$ (see appendix) and only one between them may
generate a non vanishing effect. We have calculated in eq. (\ref{MDM})
 the DM decay amplitude
for Dirac neutrinos; the
amplitude for Majorana neutrinos can be simply obtained by
doubling that formula \cite{Petcov}.
Then, from eqs.(\ref{MIB},\ref{MDM}), the interference between the inner
bremsstrahlung and the magnetic DM amplitudes is given by
\begin{eqnarray}
{\rm Re} (\overline{{\cal M}_{IB}^{\dag}{\cal M}_{DM} }) &=&
16\pi \alpha \bigl(G_F f_{\pi}\bigr)^2 m_e\sum_{ji} ^{'}\, U_{ej}U^*_{ie}\,
\kappa_{ji}\,{1\over 2(kh) - \Delta m_{ij}^2}\nonumber\\
\Biggl[ &\pm& m_i\, \left( 2{(pk)(kh) + (kh)^2 +
(ph)(kh) + m_e^2(kh)\over qk} + {2(kh)^2 - m_e^2(kh)\over pk}
\right) \nonumber\\
&+&  m_j\,\left({2(ph)(kh) +
m_e^2(kh) - m_j^2(pk)\over (qk)} - m_e^2{(kh)\over(pk)}\right) \nonumber\\
&\mp& m_im_j^2\, {(pk)\over (qk)}
\: \Biggr] \label{IBDM}
\end{eqnarray}
while from eqs. (\ref{MSD},\ref{MDM}) we obtain the interference with the
structure dependent amplitude
\begin{eqnarray}
 {\rm Re}(\overline{{\cal M}_{SD}^{\dag}{\cal M}_{DM}}) &=&
32\pi \alpha {\bigl(G_F f_{\pi}\bigr)^2 \over m_{\pi}m_e }
\sum_{ji} ^{'}\,U_{ej}U^*_{ie}\,\kappa_{ji}\, {1\over 2(kh) - \Delta m_{ij}^2}
\nonumber\\
\Biggl[ &\mp& m_i(F_V \pm F_A)\,\left(
2 m_e^2\,(kh) + m_e^2\,(pk) + 2(ph)(kh) \right)\nonumber\\
&+& m_j(F_V + F_A)\,
\bigl(m_e^2\,(kh) + m_e^2\,(pk)\nonumber\\
&+& \;m_j^2\,{(pk)^2\over (kh)} + m_j^2\,(pk) -
2{(pk)^2\over (kh)} + 2(pk)\bigr)\nonumber\\
&\pm& m_im_j^2 (F_V - F_A) (pk)\:\Biggr]    \label{SDDM}
\end{eqnarray}
where $\kappa_{ji} = {\rm Im}(\mu_{ji} + id_{ji})/\mu_B$ and $\Delta m_{ij}^2 =
m_i^2 - m_j^2$. The primes means that the sum is performed for $i \neq j$.
The oversign (lowersign)
is for pure magnetic (electric) DM process (we remember to the reader that if
$\mu_{M\,ji} \neq 0$ then CP imply $d_{M\,ji} = 0$ and vice versa).
The contribution from $|{\cal M}_{DM}|^2$ is
subdominant and we do not report it.

The interference takes place only when the chirality of the
outgoing neutrino in the DM process is the same as in the conventional
process. For this reason all the terms in (\ref{IBDM},\ref{SDDM}) are
proportional to the mass of the propagating ($m_i$) or of the
outgoing ($m_j$) neutrino.
If, as many model suggest, a mass hierarchy exist for neutrinos,
similarly to that for charged leptons, then processes with
$m_j \ll m_i$ would be
phase space favoured (see next section).
It is interesting to observe how the sign of the interference,
when the neutrino mass insertion is in the
propagator (terms proportional to $m_i$), is opposite for magnetic or
electric DM.
 We are going to discuss below as
this might supply a criterium to discriminate experimentally magnetic
from electric DM of Majorana neutrinos.
We note that, as far as we know, all the performed as proposed experiments
looking for neutrinos DM could only measure $\left|\mu + id\right|$
\cite{Raffelt2}.

\section{Kinematical distribution}

In the pion rest frame the scalar products that appears in the expressions
(\ref{IBDM},\ref{SDDM}) are:
\begin{eqnarray}
(ph) &=& {1\over 2}\,m_{\pi}^2(1 - x - r^2 - s^2)\\
(kh) &=& {1\over 2}\,m_{\pi}^2(1 - y + r^2 - s^2)\\
(pk) &=& {1\over 2}\,m_{\pi}^2(x + y - 1 - r^2 + s^2)
\end{eqnarray}
where $r = m_e/m_{\pi}$ and $s = m_j/m_{\pi}$.
The kinematical range of definition of $x$ and $y$ is
\be
y_{max,min} = 1 + r^2 - s^2 - {1\over 2}\,{x\over 1 - x}\,(1 - x - s^2 - r^2)
\pm {1\over 2}\,{x\over 1 - x}\,\sqrt{(1 - x - r^2  - s^2)^2 - 4s^2r^2}
\label{ymax}
\ee
where
\be
0 \le x \le 1 - (r + s)^2 \ .
\ee
We need to distinguish three cases: $a)$ $m_i \gg m_j$; $b)$ $m_i \ll m_j$
$c)$ $m_i \simeq m_j$. Only the case $(a)$ might be of piratical interest.
In fact,
we see from (21) that in case $(b)$ the available
phase space is reduced with respect to case $(a)$. Although this
excluded region
is not so large however, we have checked that it involves a kinematical
region where the differential decay rate takes its maximum values. The case
$(c)$ regards almost degenerate neutrino mass eigenstates.
Because we expect measurable effects only for neutrino masses in the $\MeV$
region and because the experimental limits leave open this window only for
$\tau$-neutrino, also this last case cannot have serious implications.

Concerning case $(a)$, a divergence associated to the neutrino propagator
appears when $2(kh) = \Delta m_{ij}^2 \simeq m_i^2$. This is not a physical
divergence because in such a kinematical configuration the propagating
neutrino $\nu_i$ is on-shell.
This becomes clear once one takes into account the one-loop correction
to the neutrino propagator.
Assuming $\nu_j$ to be massless we find that he
propagator has the following Breit-Wigner form:
\be
iS_{BW}(p_*) = i\,{ (\hat{p_*} + m_i)(p_*^2 - m_i^2) -
i\frac{\Gamma}{m_i}\bigl[
\hat{p_*}(p^2 - m_i^2) + 2p_*^2m_i\bigr] \over (p_*^2 - m_i^2)^2 +
4p_*^4\frac{\Gamma^2}{m_i^2}} \left( 1 + {\cal O}\left(\frac{\Gamma^2}{m_i^2}
\right)\right)
\label{BV}
\ee
where
\be
\Gamma \simeq {1\over 8\pi}\, (|\mu_{ji}|^2 + |d_{ji}|^2)\,m_i^3
\ee
is the rate for the decay $\nu_i \rightarrow \nu_j\,\gamma$, and
$p_* = k + h$.
The (\ref{BV}) is finite when the neutrino is on-shell {\it i.e.} for
$y = 1 + r^2 - (m_i/m_{\pi})^2$.
This kinematical configuration corresponds to a {\it resonance}.
It is well visible in the plot for $d\Gamma_{IB_{SD}}/dy$ (Fig. 3).
However, it can not give a relevant contribution to the total decay rate
because its width is very narrow and the positive and negative peaks
of the resonance cancel each other in the interference terms.
Finally, we observe that magnetic (electric) DM give rise to a negative
(positive) contribution to the decay rate.

The apparent discrepancy between the SM and the experimental spectra was
parametrized in ref. \cite{Bolotov} as an anomalous value for the SD$^-$
component of the spectrum. Specifically, the experimental spectrum can be
fitted if the SD$^-$ is negative and of the
same order of the SD$^+$ contribution, whereas the SM predicts a
positive SD$^-$ which is roughly one percent of the SD$^+$.
In ref. \cite{Pobla} it was shown that the interference between the
inner bremsstrahlung and a new tensorial interaction
gives rise to an additional component in the spectrum, which mimics the
SD$^-$ term and could explain the observed discrepancy if the coupling of
the tensorial term were sufficiently large \footnote{Recently it has been
advanced that such effect might be observed in the Kaons radiative decay
\cite{Gabri}}.
Nevertheless, as was pointed
out by Voloshin \cite{Volosh}, such a large coupling would induce, via
radiative corrections, an unacceptably large rate for the decay
$\pi\rightarrow e \nue$.

{}From Figs. 3 and 4, where we assume a mixing matrix element $U_{e3} =
5 \times 10^{-2}$, a neutrino mass $m_3 = 30\: \MeV$ and a neutrino magnetic
DM $\mu_{13} = 4\times 10^{-6}\: \mu_B$,
we see that the effect of neutrino DM's cannot account for
the observed discrepancy between the spectra, since it is at least two
orders of magnitude smaller than the needed value. However we note that
it might be comparable to  the SD$^-$ contribution, so that it could be
probed experimentally if the statistical errors were lowered to the one
percent level. If this effect were observed then two relevant
informations could be read off.
In the first place, massive neutrinos would be Majorana fermions, since,
as we have
discussed, in the Dirac case there is no interference with the conventional
decay amplitude. In the second place, it
would be  possible to distinguish between the electric and the magnetic DM by
looking at the sign of the non standard contribution to the spectrum
\footnote{The non-diagonal electromagnetic DM's involving the heaviest neutrino
family might be also
probed searching for the radiative decay $\pi^0 \rightarrow \gamma \nu_i
{\bar \nu_3}$ (as the CPT conjugate).
A similar process has been studied in ref. \cite{pi0} assuming
a diagonal magnetic moment for the $\nu_3$ Dirac mass eigenstate. Such
analysis apply with very slight corrections to Majorana neutrinos.
If $\nu_3$ mass is in the MeV range about a factor four is gained
in the decay rate.}.

The same conclusions reported above apply, qualitatively,
to the radiative decay
of heavier charged mesons, e.g. Kaons, although the experimental study
of these processes would involve more practical difficulties.

\vskip 1.cm
\begin{centerline}
{\bf Acknowledgments}
\end{centerline}
\vskip 0.5cm
We

We are grateful to S. Petcov and D. Zanello for useful discussions, and to
M. Lusignoli and A. Masiero for reading the manuscript and providing
helpful comments. Particularly important were conversations with
G.F. Giudice. We also thank the organizers of the 1992 Summer Institute
on Low Energy Neutrino Physics and Astrophysics, at the Gran Sasso
Laboratories, where this work was initiated.

\section{Appendix: Dirac and Majorana neutrino form factors}

{}From Lorentz invariance and current conservation, it follow that for
neutrinos, the matrix element of the electromagnetic current $J_{\mu}^{EM}$
has the general form \cite{Petcov}:
\be
\langle \nu_j|J^{EM}_{\mu}(q^2)|\nu_i\rangle = i\, \bar{u_i} \bigl[
f_{ji}(q^2)\gamma_{\mu} + g_{ji}(q^2)\, (q^2\gamma_{\mu} - 2miq_{\mu})
\gamma_5 + \mu_{ji}(q^2)\,\sigma_{\mu\nu}q^{\nu} + i\,d_{ji}(q^2)\,
\sigma_{\mu\nu}q^{\nu}\gamma_5 \bigr]\,u_i \label{A1}
\ee
where $f_{ji}(q^2), g_{ji}(q^2), \mu_{ji}(q^2)$ and $d_{ji}(q^2)$ are the form
factors and $\sigma_{\mu\nu} = {i\over 2}\,[\gamma_{\mu},\gamma_{\nu}]$.
The magnetic and electric DM, diagonal $(i=j)$ or of transition $(i
\neq j)$, are, respectively, the value of $\mu_{ji}(q^2)$ and $d_{ji}(q^2)$ at
$q^2 = 0$. Hermiticity of $J^{EM}_{\mu}$ requires
\be
\mu_{ji}^* = \mu_{ij} \qquad d_{ji}^* = d_{ij} \label{AE}
\ee
for Dirac as for Majorana  neutrinos.
However, some important differences between these kinds of neutrinos
come out when CPT and CP symmetries are imposed.

For Dirac neutrinos CPT invariance implies
\be
(\overline{\mu}_D)_{ji} = - (\mu_D)_{ij} \qquad (\overline{d}_D)_{ji} =
(d_D)_{ij}
\ee
whereas CP requires
\be
(\overline{\mu}_D)_{ji} = - (\mu_D)_{ji} \qquad (\overline{d}_D)_{ji} =
- (d_D)_{ji}\ .
\ee
Hermiticity + CPT + CP then enforce $(\mu_D)_{ji}$ to be real and symmetric
and $(d_D)_{ji}$ to be imaginary and antisymmetric. A diagonal electric
DM of neutrinos have to be vanishing if CP is conserved in the leptonic
sector but this is not necessary for off-diagonal $d_{ji}$ components.

Majorana neutrinos are defined to transform according
\be
CPT|\nu({\bf p},h)\rangle = \eta_{CPT}^h|\nu({\bf p},-h)\rangle
\ee
i.e. they are CPT self-conjugate (here $\bf{p}$ and $h$ are, respectively,
the momentum and the elicity of the neutrino). The phase factor
$\eta_{CPT}^h$ can be decomposed as $\eta_{CPT}^h = \eta_{CPT}
(-1)^{h - 1/2}$. From the Hermiticity and requiring CPT conservation
it follows \cite{Kayser}:
\be
(\mu_M)_{ji}^* = - \xi\,(\mu_M)_{ij} \qquad (d_M)_{ji}^* = - \xi\,(d_M)_{ij}
\label{AMCPT}
\ee
where $\xi = \eta_{CPT\,i}^*\eta_{CPT\,j}$. This factor does not have
any absolute significance because nothing changes replacing $|\nu_i\rangle$
with $e^{i\phi}|\nu_i\rangle$. Thus, the conditions (\ref{AE},\ref{AMCPT})
enforce $(\mu_M)_{ji}$ and $(d_M)_{ji}$ to be antisymmetric and both real
or both imaginary.

Whereas the relative CPT phase of Majorana neutrinos does not have any physical
meaning, for the relative CP phase the situation is different. Indeed, since
\be
CP|\nu(\bf{p},h)\rangle = \eta\,|\nu(-\bf{p},-h)\rangle
\ee
where $\eta = \pm i$, and $J^{EM}_{\mu} \rightarrow_{CP} J^{EM\, \dag}_{\mu}$,
if CP is conserved it follows:
\be
(\mu_M)_{ji} = + \eta_j\eta_i\,(\mu_M)_{ji} \qquad
(d_M)_{ji} = - \eta_j\eta_i\,(d_M)_{ji}\ .
\ee
This means that when $(\mu_M)_{ji} \neq 0$ (thus, $(d_M)_{ji} = 0$) the
electromagnetic vertex connects neutrino states with opposite CP parities
only, whereas when $(d_M)_{ji} \neq 0$ (thus, $(\mu_M)_{ji} = 0$) it connects
states with the same  CP parities.

\vskip 1.cm
\noindent{\bf Figure Caption}
\begin{itemize}
\item[{\bf Fig. 1}]{ Feynmann diagram for the process $\pi^- \rightarrow
e^- \nue \gamma$ when the photon is emitted from the neutrino leg.}
\item[{\bf Fig. 2}]{ Loop diagram for the lowest order contribution to
the radiative pion decay through neutrino non-diagonal DM (see text)}
\item[{\bf Fig. 3a}]{ Contributions to the differential decay rate $d\Gamma/dx$
integrated
for $1 - 0.8 x \le y \le 1 + r^2$. The outgoing neutrino is assumed to be
massless, whereas the propagating one has a mass of 30 MeV. Numerical
values for $d\Gamma/dx$ are arbitrary. Note that
$SD-DM$ interference term is taken with absolute value.}
\item[{\bf Fig. 3b}]{ The same as in Fig. 3a for $d\Gamma/dy$ integrated
for $0.3 \le x \le 1 - r^2$}.
\end{itemize}

\end{document}